\documentclass[twocolumn,showpacs,preprintnumbers,amsmath,amssymb]{revtex4}

\usepackage{graphicx}
\usepackage{dcolumn}
\usepackage{bm}
\usepackage{latexsym}

\begin{document}

\preprint{APS/paper}

\title{Phenomenological theory of phase transitions in epitaxial Ba$_x$Sr$_{1-x}$TiO$_3$ thin films}

\author{ V. B. Shirokov }
\author{ Yu. I. Yuzyuk }
\affiliation{Department of Physics, Southern Federal University, Zorge 5, Rostov-on-Don, 344090, Russia}
\author{ B. Dkhil }
\affiliation{Laboratoire Structures Propri\'et\'es et Mod\'elisation des Solides, Ecole Centrale Paris, UMR-CNRS 8580, F-92290 Chatenay-Malaby, France}
\author{ V. V. Lemanov }
\affiliation{A.F. Ioffe Physico-Technical Institute, 194021 St. Petersburg, Russia}

\date{\today}

\begin{abstract}
A phenomenological thermodynamic theory of Ba$_x$Sr$_{1-x}$TiO$_3$ (BST-$x$) thin films epitaxially grown on cubic substrates is developed using the Landau-Devonshire approach. The eighth-order thermodynamic potential for BT single crystal  and modified fourth-order potential for ST single crystal were used as starting potentials for the end-members of the solid solution with the aim to develop potential of BST-$x$ solid solution valid at high temperatures. Several coefficients of these potentials for BT were changed to obtain reasonable agreement between theory and experimental phase diagram for BST-$x$ ($x > 0.2$) solid solutions. For low Ba content we constructed the specific phase diagram where five phases converge at the multiphase point ($T_{N2} = 48$ K, $x = 0.029$) and all transitions are of the second order. The "concentration-misfit strain"  phase diagrams for BST-$x$ thin films at room temperature and "temperature-misfit strain"  phase diagrams for particular concentrations are constructed and discussed. Near $T_{N2}$ coupling between polarization and structural order parameter in the epitaxial film is modified considerably and large number of new phases not present in the bulk materials appear on the phase diagram.
\end{abstract}


\maketitle

\section{\label{sec:level1}INTRODUCTION}

Ferroelectric thin films are very good candidates for a wide range of applications: for example, as high-density dynamic random access memories, large-scale integrated capacitors, pyroelectric detectors, micro- and nanoelectromechanic devices \cite{c1,c2}. Epitaxially grown thin films are usually highly constrained because their fabrication is accompanied by several strain factors originating due to lattice mismatch and the difference between the thermal-expansion coefficients of the film and the substrate. Also, transformation strain usually appears at the ferroelectric phase transition if the heterostructure deposed at elevated temperatures is cooled below the Curie temperature ($T_c$).The importance of strain effects in determining the properties of thin-film ferroelectrics was recognized in numerous theoretical and experimental investigations \cite{c3,c4,c5,c6,c7,c8,c9,c10,c11,c12,c13}.

The Ba$_x$Sr$_{1-x}$TiO$_3$ (BST-$x$) solid solution is one of the most studied lead-free ferroelectric system from fundamental and technological point of view. In the paraelectric cubic phase, the end-members BaTiO$_3$ (BT) and SrTiO$_3$ (ST) have similar crystal structure but exhibit quite different phase transition sequences on cooling. BT displays three phase transitions towards ferroelectric phases characterized by the polarization along different directions. On the other side, ST shows no ferroelectric phase transition on cooling but first an antiferrodistorsive transition followed by the appearance, at very low temperature, of a quantum paraelectric state. Both BT and ST compounds have been widely investigated, both theoretically and experimentally, whereas, the efforts on the BST-$x$ solid solution are mainly experimental. It is then very desirable to develop new calculation and theoretical tools that allow better and deeply describe the structure and thus the properties of this system. The temperature-concentration phase diagram of bulk BST-$x$ solid solutions based on the phenomenological six-order thermodynamic potential was recently developed \cite{c14} and compared with available experimental data.

Thin films of BST-$x$ solid solutions are of great technological interest due to their excellent ferroelectric and piezoelectric properties. The structure and properties of ferroelectric thin films substantially differ from those of bulk ferroelectrics; therefore, theoretical analysis is important since it can provide fundamental insights into the behavior of thin films. Recently, Ban and Alpay \cite{c15} have developed phase diagrams for single-domain epitaxial Ba$_{0.6}$Sr$_{0.4}$TiO$_3$ and Ba$_{0.7}$Sr$_{0.3}$TiO$_3$ films on (001) cubic substrates as a function of the misfit strain based on the Landau-Devonshire six-order potential as a polynomial of the polarization components in accordance with the one developed by Pertsev $et$ $al$ \cite{c5}. The parameters used for the calculation of the renormalized coefficients for BST films were obtained by simple averaging the corresponding parameters of BT and ST. It is worth noting that the contribution of six-order polarization terms to the free energy was neglected in their calculations that is rather oversimplified approach. However, as known from experimental investigations, high-order terms should be taken into account, because the two-dimensional clamping increases considerably the ferroelectric phase transition temperature in perovskite thin films. Stress-induced upward shift in $T_c$ as large as tens of degrees have been observed \cite{c9,c11}. In this case, the coefficients at the sixth-order terms in the thermodynamic potential \cite{c5} becomes negative, therefore, the six-order expansion is not valid for high-temperature phase transitions in thin films. Very recently \cite{c16}, the eighth-order Landau-Devonshire potential was used to construct "misfit-temperature" phase diagrams of epitaxial BT thin films on cubic substrates.

The present paper is devoted to BST-$x$ thin films and is organized as follows. First, we develop thermodynamic potential for BST-$x$ solid solutions using phenomenological models known for pure BT and ST single crystals. Due to the above mentioned reason the sixth-order potential for BST solid solutions used in \cite{c14} is not valid for thin films, therefore here we use eighth-order potential developed by Li \cite{c17} for BT single crystal and fourth-order potential recently developed for ST crystal \cite{c18}. It is worth noting that in \cite{c14} the coefficient at $p^2$ in the potential for ST was changed considerably to achieve agreement between experimental data \cite{c19} and para-ferroelectric transition line on the theoretical phase diagram. As a result, the value of the susceptibility was found to be significantly and abnormally large with respect to the experimentally observed one. In the present work we use coefficients of thermodynamic potential for ST \cite{c18} and BT \cite{c17}. Moreover, several coefficients (Q$_{11}$, Q$_{12}$,  $\alpha_{123}$) from \cite{c17} were changed to get better agreement with experimental diagram in the Ba-rich side. Second, we develop thermodynamic potential for BST thin film epitaxially grown on cubic substrate. The potential derived for BST solid solutions according to the method proposed in \cite{c14} includes no term associated with thermal expansion. To take into account thermal expansion the above method was modified with the aim to include evidently relevant term in the potential. Finally, the resulting potential was used to construct according to the conventional way \cite{c5} the "misfit-temperature" diagrams for thin films of particular BST-$x$ compositions.

\section{A PHENOMENOLOGICAL THERMODYNAMIC POTENTIAL FOR BST SOLID SOLUTION}

Following the method developed in \cite{c14} the Helmholtz thermodynamic potential of the solid solution can be written using the known thermodynamic potentials $F_{ST}(\eta,u)$ and $F_{BT}(\eta,u)$ for end members of the solid solution $x = 0$ (ST) and $x = 1$ (BT):
\begin{equation}
F = (1 - x)F_{ST} (\eta ,u - \Delta _{ST} ) + xF_{BT} (\eta ,u - \Delta _{BT} )
\end{equation}
where  $\eta$ is the order parameter,$u$ is the common elastic strain of the solid solution, $\Delta _{ST}$ and $\Delta _{BT}$  are lattice strains of the end members, needed to fit the lattice parameter ST, BT and BST-$x$ \cite{c14}. The lattice parameters of the solid solution end-members are
\begin{equation}
\begin{gathered}
  a_{ST}  = a_x (1 + \Delta _{ST} ), \hfill \\
  a_{BT}  = a_x (1 + \Delta _{BT} ). \hfill \\
\end{gathered}
\end{equation}
The lattice parameter of the solid solution $a_x$ can be found from the condition, which implies complete compensation of internal elastic forces:
\begin{equation}
\begin{array}{l}
(1 - x)\left. {\frac{{\partial F_{ST} \left( {\eta ,u - \Delta _{ST} } \right)}}
{{\partial u}}} \right|_{\eta ,u = 0}
 + x\left. {\frac{{\partial F_{BT} \left( {\eta ,u - \Delta _{BT} } \right)}}
{{\partial u}}} \right|_{\eta ,u = 0}  = 0
\end{array}
\end{equation}

The thermodynamic description may be developed starting from the power-series expansion of the Gibbs potential $\Phi(\eta,t)$. The relevant general expression for cubic perovskite is given as

\begin{equation}
\begin{array}{l}
\Phi  = \beta _1 \left( {\varphi _1^2  + \varphi _2^2  + \varphi _3^2 } \right) + \beta _{11} \left( {\varphi _1^4  + \varphi _2^4  + \varphi _3^4 } \right) \\
\qquad  + \beta _{12} \left( {\varphi _1^2 \varphi _2^2  + \varphi _1^2 \varphi _3^2  + \varphi _2^2 \varphi _3^2 } \right)   + \alpha _1 \left( {p_1^2  + p_2^2  + p_3^2 } \right) \\
\qquad  + \alpha _{11} \left( {p_1^4  + p_2^4  + p_3^4 } \right) + \alpha _{12} \left( {p_1^2 p_2^2  + p_1^2 p_3^2  + p_2^2 p_3^2 } \right)  \\
\qquad    - t_{11} \left( {\varphi _1^2 p_1^2  + \varphi _2^2 p_2^2  + \varphi _3^2 p_3^2 } \right) - t_{12} [\varphi _1^2 (p_2^2  + p_3^2 ) \\
\qquad    + \varphi _2^2 (p_1^2  + p_3^2 ) + \varphi _3^2 (p_1^2  + p_2^2 )] - t_{44} (\varphi _2 \varphi _3 p_2 p_3  \\
\qquad    + \varphi _1 \varphi _3 p_1 p_3  + \varphi _1 \varphi _2 p_1 p_2 )   + \Phi _6  + \Phi _8  + \Phi _t   \\
\end{array}
\end{equation}

where $p$ is the order parameter - polarization related to ionic shifts in polar zone-center $F_{1u}$ mode,  $\varphi$ is the out-of-phase rotation of TiO$_6$ octahedra corresponding to the $R_{25}$ zone-boundary mode in the cubic phase $Pm$3$m$(O$_h^1$).

The high-order terms are written as
\begin{equation}
\begin{array}{l}
\Phi _6  = \alpha _{111} \left( {p_1^6  + p_2^6  + p_3^6 } \right) + \alpha _{112} [p_1^4 (p_2^2  + p_3^2 )  + p_2^4 (p_1^2  + p_3^2 )\\
\qquad  + p_3^4 (p_1^2  + p_2^2 )] + \alpha _{123} p_1^2 p_2^2 p_3^2 ,  \\
\Phi _8  = \alpha _{1111} (p_1^8  + p_2^8  + p_3^8 ) + \alpha _{1112} [p_1^6 (p_2^2  + p_3^2 ) + p_2^6 (p_1^2  + p_3^2 ) \\
\qquad  + p_3^6 (p_1^2  + p_2^2 )]  + \alpha _{1122} (p_1^4 p_2^4  + p_1^4 p_3^4  + p_2^4 p_3^4 ) \\
\qquad  + \alpha _{1123} (p_1^4 p_2^2 p_3^2  + p_1^2 p_2^4 p_3^2  + p_1^2 p_2^2 p_3^4 )  \\
\end{array}
\end{equation}

The elastic energy $\Phi _t$ is:

\begin{equation}
\begin{array}{l}
\Phi _t  =  - R_{11} \left( {t_1 \varphi _1^2  + t_2 \varphi _2^2  + t_3 \varphi _3^2 } \right) - R_{44} (t_4 \varphi _2 \varphi _3  + t_5 \varphi _1 \varphi _3 \\
\qquad    + t_6 \varphi _1 \varphi _2 )   - R_{12} [t_1 (\varphi _2^2  + \varphi _3^2 ) + t_2 (\varphi _1^2  + \varphi _3^2 )  \\
\qquad  + t_3 (\varphi _1^2  + \varphi _2^2 )] - Q_{11} \left( {t_1 p_1^2  + t_2 p_2^2  + t_3 p_3^2 } \right) \\
\qquad  - Q_{44} \left( {t_4 p_2 p_3  + t_5 p_1 p_3  + t_6 p_1 p_2 } \right)  \\
\qquad    - Q_{12} \left( {t_1 (p_2^2  + p_3^2 ) + t_2 (p_1^2  + p_3^2 ) + t_3 (p_1^2  + p_2^2 )} \right)  \\
\qquad  - \frac{1}{2}s_{11} \left( {t_1^2  + t_2^2  + t_3^2 } \right) - \frac{1}{2}s_{44} \left( {t_4^2  + t_5^2  + t_6^2 } \right)  \\
\qquad - s_{12} \left( {t_1 t_2  + t_1 t_3  + t_2 t_3 } \right) - \lambda T\left( {t_1  + t_2  + t_3 } \right)  \\
\end{array}
\end{equation}

where $s_{kj}$ are the compliances,  $\lambda$ is the linear thermal expansion coefficient, $T$ is the absolute temperature, the stresses $t_i$, i=1..6 are given in Voigt notations. In the following, the coefficients of the potentials (4)-(6) of the end-members of the solid solution are denoted by additional indexes ST and BT. The reference point on the temperature scale is taken at $T = 0$. Temperature $T$ should be substituted in Eq. 6 by  $\Delta T=T-T_0$ if reference point $T_0$ is not equal to zero.

Eq. (1) is written for Helmholtz potential $F(\eta,u)$ , which can be obtained from Gibbs potential (4), (5), (6) by the formal substitution of the coefficients and substitution of stress $t$ in (4) for the quantities determined from the equations $u_i  =  - \frac{{\partial \Phi }}{{\partial t_i }}$  \cite{c20}. Only coefficients at second and fourth order terms in (4) will be renormalized because Eq. (6) contains quadratic strains.

The Helmholtz potential of the solid solution (1) contains no distinct term corresponding to the thermal expansion. Actually, temperature dependence of the lattice parameters is included in the lattice parameter of the solid solution $a_x$. To construct the potential in a conventional form, when linear thermal expansion is ascribed by the terms linear with respect to strains, one has to shift the common strain $u$ by the value of the linear thermal expansion. The latter can be found from the linear expansion of $a_x$ (Eq.7 in Ref.14). However we used the following way. In Eq.3 we exclude the terms corresponding to thermal expansion, therefore we assume $\Delta_{ST}$ and $\Delta_{BT}$  to be temperature independent. In this case, $\Delta_{ST}$  and $\Delta_{BT}$  derived from Eqs. (2)- (3) preserve the term corresponding to the thermal expansion in Eq. (1). Under the above introduced conditions, in the cubic paraelectric phase Eqs. (2) and (3) yield:
\begin{equation}
\begin{array}{l}
  \Delta _{ST}  = \frac{{ - x\delta }}
{{(1 - x)\tau  + (1 + \delta )x}}, \hfill \\
  \Delta _{BT}  = \frac{{(1 - x)\tau \delta }}
{{(1 - x)\tau  + (1 + \delta )x}}, \hfill \\
  a_x  = \frac{{\left( {1 - x} \right)\tau a_{ST}  + xa_{BT} }}
{{\left( {1 - x} \right)\tau  + x}} \hfill \\
\end{array}
\end{equation}

The thermal expansion coefficient can be written as:
\begin{equation}
\lambda _x  = \frac{{\left( {\left( {1 - x} \right)\tau  + \left( {1 + \gamma } \right)x} \right)}}
{{\left( {1 - x} \right)\tau  + x}}\lambda _{ST}
\end{equation}
where $\tau  = \frac{{s_{11,BT}  + 2s_{12,BT} }} {{s_{11,ST}  + 2s_{12,ST} }}, \gamma  = \frac{{\lambda _{BT}  - \lambda _{ST} }} {{\lambda _{ST} }}, \delta  = \frac{{a_{BT}  - a_{ST} }} {{a_{ST} }} $. Eq. (8) can be easily obtained from Eq.7 of Ref. 14 if the lattice parameter of the solid solution is presented as one-series expansion $ a_{BST}  = a_x (1 + \lambda _x T)$.

\section{PHASE DIAGRAM OF THE BST-$x$ SOLID SOLUTION}

The starting Gibbs potential for pure BT is expanded as polynomial of polarization components up to eight order \cite{c17} and contains only additional $\varphi^2$ term with the positive coefficient equal to  $3.7 \times 10^{29}$ $J/m^5$ Ref. 14. All coefficients are listed in Table I, where coefficients in Eqs. (4)- (6) in higher-order terms containing  $\varphi$ are equal to zero. Note, the coefficient $\alpha_{123}$  used in Ref. 17 was changed in our calculations to fit theoretical orthorhombic-rhombohedral phase transition temperature to the experimental value. All coefficients in Gibbs potentials for end-members BT and ST used in this work are listed in Table I.

It is important to emphasize that it is not easy to deal with the coefficients especially in the Sr-rich region where both the quantum effects and the coupling between ferroelectricity and antiferrodistorsivity are coexisting. As a consequence, the coefficients of the potential depend strongly on the technological parameters of the samples synthesis including homogeneity, impurities, vacancies, stress conditions, grain boundaries, but also on the techniques (Raman spectroscopy, dielectric spectroscopy, x-ray diffraction, $etc$) used to have access to these coefficients. As known, even dielectric properties are very sensitive to the conditions of the sample preparation \cite{c22}, thus coefficients in the $p^2$ terms depend on these parameters too. Also, as follows from the proposed method of derivation of thermodynamic potential of solid solution, any impurities induce additional deformations, which renormalize coefficients of the potential. As a consequence, coordinates of critical points can be different for samples in single crystal, powder or ceramic form prepared in different laboratories. Nevertheless, even if the critical points may differ from one sample to another one, the qualitative phase diagram should not change and can serve as a starting setting to study the consequences when BST-$x$ is as thin film form.

\begin{table}
\caption{\label{tab1}
Coefficients of Gibbs potentials for BT \cite{c17} and ST \cite{c18}. Coefficients in the quadratic terms are: $\beta _{\text{1,ST}} = 1.036 \times 10^{28} \text{[coth(43.8} /T \text{) - coth(43.8/106)] }$, $\beta _{\text{1,BT}} = 3.7 \times \text{10}^{29} $ in J/m$^5$, $\alpha _{\text{1,ST}}  = 4.05 \times 10^7  \times \text{[coth(54/} T \times{) - coth(54/30)]}$, $\alpha _{\text{1,BT}} = 4.124 \times 10^5 \text{(} T \text{ - 388)}$ in Jm/C$^2$. Coefficients from Refs. \cite{c17,c21} different from those used here are given in brackets.
}
\begin{ruledtabular}
\begin{tabular}{cccc}
Coefficient& SrTiO$_3$ & BaTiO$_3$ & Units  \\ \hline \\

$\beta_{11}$  &  1.69  &  0  &  $\times 10^{50}$ J/m$^7$\\
$\beta_{12}$  &  4.07  &  0  & \\

$\alpha_{11}$ &  1.04 (17)    &  -2.097  & $\times 10^{8}$ Jm$^5$/C$^4$\\
$\alpha_{12}$ &  0.746 (13.7)  &  7.974   & \\

$\alpha_{111}$ &  0  &  1.294      &  $\times 10^{9}$ Jm$^9$/C$^6$\\
$\alpha_{112}$ &  0  &  -1.950     &  \\
$\alpha_{123}$ &  0  &  -0.76(-2.5)&  \\

$\alpha_{1111}$ & 0  &  3.863  &  $\times 10^{10}$ Jm$^{13}$/C$^8$\\
$\alpha_{1112}$ & 0  &  2.529  &  \\
$\alpha_{1122}$ & 0  &  1.637  &  \\
$\alpha_{1123}$ & 0  &  1.367  &  \\

$t_{11}$  &  -1.74      &  0  &  $\times 10^{29}$ J/C$^2$m\\
$t_{12}$  &  -0.75      &  0  &  \\
$t_{44}$  &  0.1(5.85) &  0  &  \\

$R_{11}$  &  0.87&0   &  $\times 10^{19}$ m$^{-2}$\\
$R_{12}$  &  -0.78&0  &  \\
$R_{44}$  &  -1.84&0  &  \\

$Q_{11}$  &  4.96(4.57)  &  11    & $\times 10^{-2}$ m$^4$/C$^2$\\
$Q_{12}$  &  -1.31       &  -4.5  & \\
$Q_{44}$  &  1.9       &  2.9   & \\

$s_{11}$  &  3.52   &  8.33   &  $\times 10^{-12}$ m$^3$/J\\
$s_{12}$  &  -0.85  &  -2.68  &  \\
$s_{44}$  &  7.87   &  9.24   &  \\

\end{tabular}
\end{ruledtabular}
\end{table}

\begin{figure}
\includegraphics{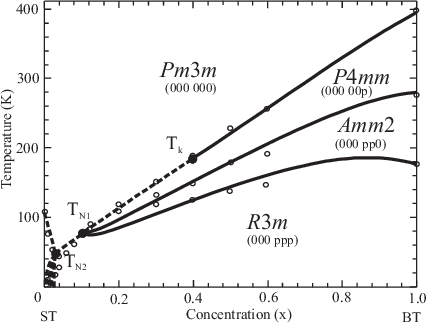}
\caption{\label{fig1} Phase diagram of bulk BST-$x$ solid solutions. Solid lines and dashed lines correspond to first and second order phase transitions, respectively. Experimental points (circles) taken from Ref. 19. The calculated coordinates of the tricritical and multicritical points are $T_{N1}= 116$ K$, x = 0.2$; $T_{N2} = 48$ K$, x = 0.029$; $T_k =317$ K$, x = 0.75$. Detailed diagram at low $x$ is presented in Fig. 2 }
\end{figure}

The resulting phase diagram for BST-$x$ solid solutions calculated using the coefficients listed in Table I is presented in Fig. 1. Although phase diagram has very similar overall view with respect to that previously reported \cite{c14}, there are some changes caused by the above mentioned changes of the coefficients in fourth-order terms: 1). The $N$-phase point $T_{N1}$ is now shifted from $x = 0.13$ to $x = 0.11$. 2). The tricritical point now appears at $x = 0.37$. 3). At low $x$ five possible diagrams were proposed in \cite{c14} for different possible $t_{44}$ values. Since $t_{44}$ coefficient was determined \cite{c18} from the experimental data, only one diagram at low $x$ values is calculated (see in Fig. 2).

\begin{figure}
\includegraphics{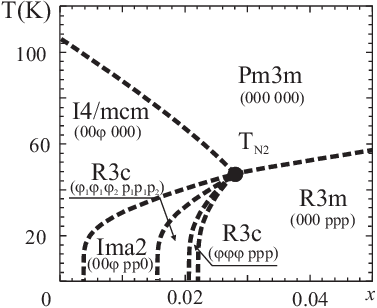}
\caption{\label{fig2}
Calculated phase diagram of bulk BST-$x$ in the vicinity of multiphase point $T_{N2} = 48$ K$, x = 0.029$. Dashed lines correspond to second order phase transition }
\end{figure}

\section{PHASE DIAGRAMS OF THIN FILMS}

Below we consider BST-$x$ thin films epitaxially grown onto (001) surface of a cubic substrate. Deposition usually occurs well above the $T_c$ of bulk BST-$x$. The film is constrained due to lattice mismatch between the film and the substrate. This deformation occurs in the plane parallel to the substrate and induces symmetry lowering from cubic $Pm3m$ to tetragonal $P4/mmm$. As a consequence, all three-dimensional irreducible representations of the cubic phase split into one- and two-dimensional representations. The group-theoretical analysis of the structures derived from the aristotype cubic perovskite $Pm3m$ by two order parameters ($F_{1u}$ and $R_{25}$ soft modes) yields 26 low-symmetry phases, while symmetry lowering of the paraelectric to tetragonal phase induces 33 phases. A list of possible low-symmetry phases derived though TiO$_6$ octahedral tilting and Ti-cation displacements is given in Table II together with their notations already used in previous literature \cite{c5, c23, c24}.
\begin{figure}
\includegraphics{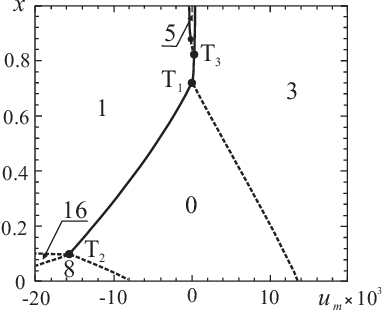}
\caption{\label{fig3}
The "concentration-misfit strain" diagram for BST-$x$ thin films at $T=300$K. The coordinates of the multiphase points are $T_1 (u_m= 0, x=0.7)$, $T_2 (u_m= -15.5 \times 10^{3}, x=0.083)$ and $T_3 (u_m=0.26 \times 10^{3}, x=0.81)$. Dashed lines correspond to the second-order phase transitions. Phase numbering is given in Table II. }
\end{figure}

In the absence of external forces the strains at the film substrate interface are $u_{11}  = u_{22}  = u_m$, $u_{12}  = 0$, where the misfit strain  $u_m$ exhibits temperature dependence \cite{c16}:
\[
u_m  = u_0  + (\lambda  - \frac{{a_S }} {a}\lambda _S )\left( {T - T_0 } \right)
\]
where $u_0  \approx \frac{{s_{11}  + s_{12} }} {{s_{11}  - s_{12} }}\frac{{a - c}} {c} $ is the primary deformation of the film at the deposition temperature $T_0$, $a$ and $c$ - the corresponding in-plane and out-of-plane lattice parameters of the film at $T_0$, $a_S$ is the lattice parameter of the substrate, $\lambda$ and $\lambda_S$ - the thermal expansion coefficients of the film and the substrate, respectively.
\begin{table}
\caption{\label{tab2}
 Low-symmetry phases allowed for BST-$x$ thin film deposited on (001) cubic substrate. For all phases with nonzero order parameter $\varphi$  the primitive cell volume is doubled as compareed to that in the high-symmetry phase. The $\varphi_i$ and $p_i$ components correspond to $R_{25}$ and $F_{1u}$ soft mode, respectively }
\begin{ruledtabular}
\begin{tabular}{lllll}

Phase    & Order                                        &   Symmetry   &  \multicolumn{2}{c}{ Notations} \\
         & parameter                                    &              &   \multicolumn{2}{c}{from Refs.} \\
         &  \bm{$\varphi \oplus p$}                          &              &  [5, 24]  &  [17]  \\ \hline \\
\textbf{0}   &  $(000 \text{ } 000)$                          &  $D_{4h}^1=P4/mmm(N123)$      &  HT       &  TP  \\
\textbf{1}   &  $(000 \text{ } 00p)$                          &  $C_{4v}^1=P4mm(N99)$         &  FTI, c   &  TF1 \\
\textbf{2}   &  $(000 \text{ } 0p0)$                          &  $C_{2v}^1=Pmm2(N25)$         &  a        &  OF1 \\
\textbf{3}   &  $(000 \text{ } pp0)$                          &  $C_{2v}^{14}=Amm2(N38)$      &  FOI, aa  &  OF2 \\
\textbf{4}   &  $(000 \text{ } p_1p_1p_2)$                    &  $C_s^3=Cm(N8)$               &  r        &  \\
\textbf{5}   &  $(000 \text{ } p_10p_2)$                      &  $C_s^1=Pm(N6)$               &  ac       &  \\
\textbf{6}   &  $(000 \text{ } p_1p_20)$                      &  $C_s^1=Pm(N6)$               &           & \\
\textbf{7}   &  $(000 \text{ } p_1p_2p_3)$                    &  $C_1^1=P1(N1)$               &           & \\
\textbf{8}   &  $(00\varphi  \text{ } 000)$                      &  $D_{4h}^{18}=I4/mcm(N140)$   &  ST       &  TS  \\
\textbf{9}   &  $(\varphi00 \text{ } 000)$                       &  $D_{2h}^{23}=Fmmm(N69)$      &  SO       &  OS1  \\
\textbf{10}  &  $(\varphi\varphi 0 \text{ } 000)$                   &  $D_{2h}^{28}=Imcm(N74)$      &           &  OS2  \\
\textbf{11}  &  $( \varphi_1 \varphi_1 \varphi_2 \text{ } 000)$        &  $C_{2h}^6=C2/c(N15)$         &           &  \\
\textbf{12}  &  $( \varphi_1 0 \varphi_2 \text{ } 000)$             &  $C_{2h}^3=C2/m(N12)$         &           &  \\
\textbf{13}  &  $( \varphi_1 \varphi_20 \text{ } 000)$              &  $C_{2h}^3=C2/m(N12)$         &           &  \\
\textbf{14}  &  $( \varphi_1 \varphi_2 \varphi_3 \text{ } 000)$        &  $C_i^1=Pi(N2)$               &           &  \\
\textbf{15}  &  $(\varphi\varphi  0 \text{ } pp0)$                  &  $C_{2v}^{22}=Ima2(N46)$      &  FOIV     &  OF6  \\
\textbf{16}  &  $(00 \varphi \text{ } 00p)$                      &  $C_{4v}^{10}=I4cm(N108)$     &  FTII     &  TF2  \\
\textbf{17}  &  $(\varphi\varphi  0 \text{ } p\text{--}p0)$         &  $C_{2v}^{20}=Imm2(N44)$      &  FOIV     &  \\
\textbf{18}  &  $(00\varphi  \text{ } pp0)$                      &  $C_{2v}^{22}=Ima2(N46)$      &  FOIII    &  OF5  \\
\textbf{19}  &  $(\varphi 00 \text{ } p00)$                      &  $C_{2v}^{18}=Fmm2(N42)$      &           &  \\
\textbf{20 } &  $(00\varphi  \text{ } 0p0)$                      &  $C_{2v}^{18}=Fmm2(N42)$      &           &  OF4  \\
\textbf{21}  &  $(\varphi 00 \text{ } 0p0)$                      &  $C_{2v}^{18}=Fmm2(N42)$      &  FOII     &  OF3  \\
\textbf{22 } &  $(\varphi 00 \text{ } 00p)$                      &  $C_{2v}^{18}=Fmm2(N42)$      &           &  \\
\textbf{23}  &  $(\varphi\varphi  0 \text{ } 00p)$                  &  $C_{2v}^{22}=Ima2(N46)$      &           &  \\
\textbf{24}  &  $(00 \varphi \text{ } p_1p_20)$                  &  $C_s^3=Cm(N8)$               &           &  \\
\textbf{25}  &  $(0 \varphi_1 \varphi_2 \text{ } p00)$              &  $C_2^3=C2(N5)$               &           &  \\
\textbf{26}  &  $(\varphi \text{--}\varphi 0 \text{ } p_1p_1p_2)$   &  $C_s^3=Cm(N8)$               &           &  \\
\textbf{27}  &  $(\varphi 00 \text{ } 0p_1p_2)$                  &  $C_s^3=Cm(N8)$               &           &  \\
\textbf{28}  &  $( \varphi_1 \varphi_20 \text{ } 00p)$              &  $C_2^3=C2(N5)$               &           &  \\
\textbf{29}  &  $( \varphi_1 \varphi_1 \varphi_2 \text{ } p\text{--}p0)$&  $C_2^3=C2(N5)$               &           &  \\
\textbf{30}  &  $( \varphi_1 \varphi_1 \varphi_2 \text{ } p_1p_1p_2)$  &  $C_s^4= Cc(N9)$              &           &  \\
\textbf{31}  &  $(0 \varphi1 \varphi_2 \text{ } 0p_1p_2)$          &  $C_s^3=Cm(N8)$               &           &  \\
\textbf{32}  &  $( \varphi_1 \varphi_20 \text{ } p_1p_20)$          &  $C_s^3=Cm(N8)$               &           &  \\
\textbf{33}  &  $( \varphi_1 \varphi_2 \varphi_3 \text{ } p_1p_2p_3)$  &  $C_1^1=P1(N1)$               &           &  \\

\end{tabular}
\end{ruledtabular}
\end{table}

\begin{figure*}
\includegraphics{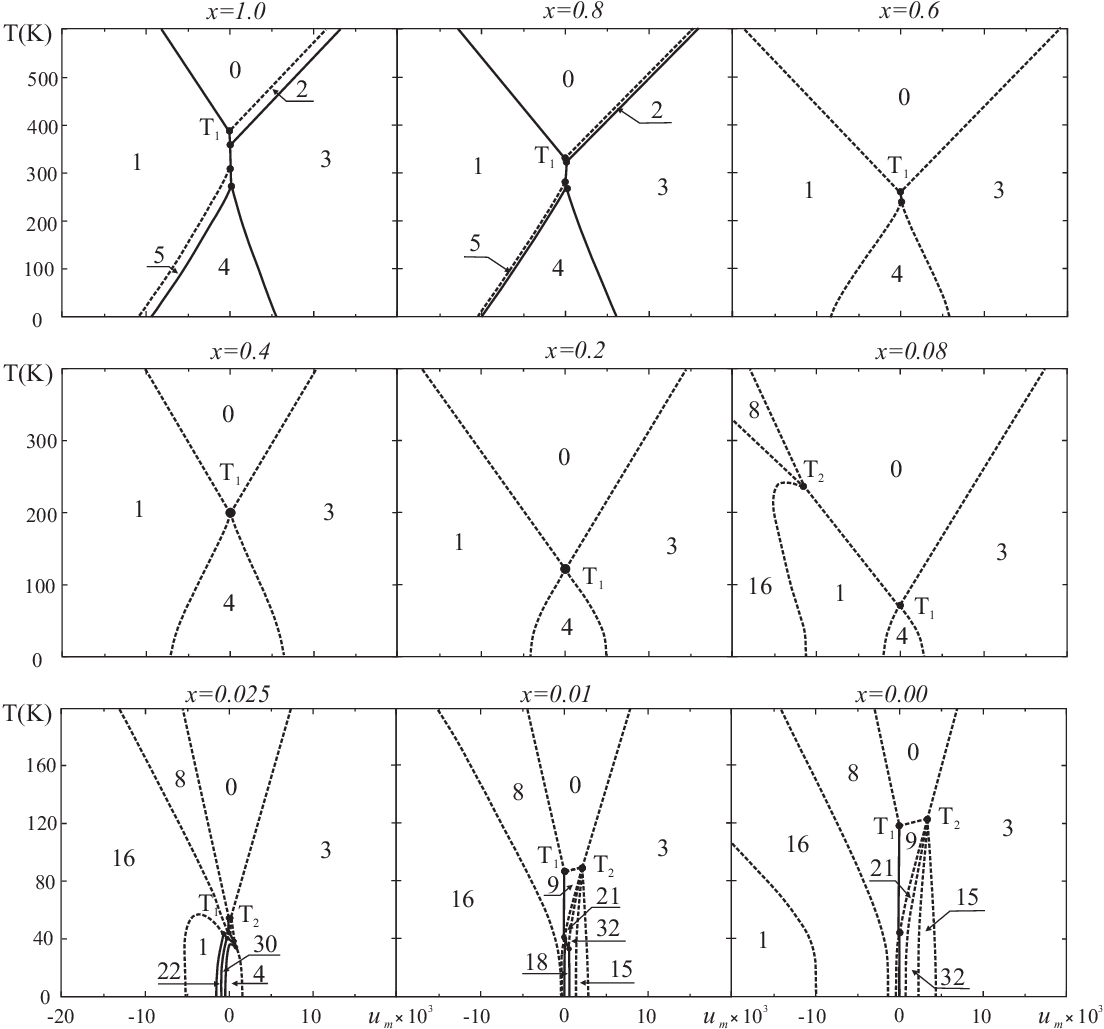}
\caption{\label{fig4}
$u_m-T$  phase diagrams of BST-$x$ thin films for selected concentrations. Solid lines and dashed lines correspond to first and second order phase transitions, respectively. Phase numbering is given in Table II. Detailed diagrams at low $x$ around $u_m = 0$ is presented in Fig. 5. }
\end{figure*}

\begin{figure}
\includegraphics{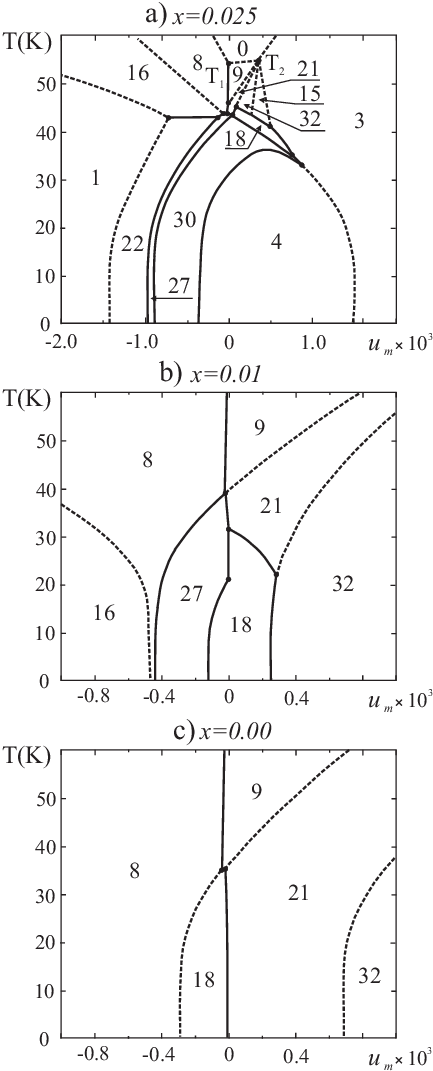}
\caption{\label{fig5}
Phase diagrams of BST-$x$ thin films at low $x$ around $u_m$ = 0. Solid lines and dashed lines correspond to first and second order phase transitions, respectively. Phase numbering is given in Table II. }
\end{figure}

Following Pertsev $et$  $al$ \cite{c5} we use the Gibbs thermodynamic potential (4) with the coefficients listed in Table I to construct the "concentration-misfit strain" ($x-u_m$ ) phase diagrams for BST thin films at room temperature and "temperature-misfit strain" ($T-u_m$) phase diagrams for particular concentrations. Recently \cite{c16, c25}, we have illustrated that phase diagrams for epitaxial BT films depend on the values of compliances and electrostrictive coefficients used in calculations. In the present work we performed calculations using the set of coefficients suitable for the whole phase diagram of BST-$x$ solid solutions.

The $x-u_m$  phase diagram for BST-$x$ thin films at $300$ K is shown in Fig. 3. There are five low-symmetry phases and two multiphase points $T_1$: $(u_m=0, x=0.72)$, $T_2$:  $(u_m= -15.7 \times{10^3}, x=0.1)$ and $T_3$:  $(u_m= 0.26 \times{10^3}, x=0.81)$, which is very sensitive to the $Q_{ij}$ values. The tetragonal paraelectric phase \textbf{0} may be stable at room temperature up to $x = 0.72$ at zero misfit strain. The monoclinic phase \textbf{5} - ($p_1 \ne 0$, $p_2 = 0$ and $p_3 \ne 0$) , with the polarization tilted with respect to the film surface is stable above $T_3$ point and at low misfit strains $u_m$. The tetragonal phase \textbf{1} with the polarization normal to the substrate ($p_1 = p_2 = 0$ and $p_3 \ne 0$) is stable in a wide range of negative misfit strains and $x > 0.1$. In the right side of the diagram the orthorhombic phase \textbf{3} ( $p_1 = p_2 \ne 0$ and $p_3 = 0$) is stable. These phases were already discussed in previous literature \cite{c15} for BST-$x$ thin films. As follows from Fig. 3, new phases may appear at room temperature only at high negative misfit strains and only in the Sr-rich side of the phase diagram, where the two-dimensional (2D) clamping (negative $u_m$) stabilizes the $I4/mcm$ tetragonal phase \textbf{8}, which exists in bulk ST crystals below $106$ K. Finally, very strong 2D clamping stabilizes tetragonal $I4cm$ polar phase \textbf{16} in the narrow concentration interval in the vicinity $x =0.1$.

Three-dimensional $u_m-x-T$ phase diagram is rather complicated and contains the multiphase point $T_N ( u_m= 0, x = 0.029, T = 48$ K  where all second-order terms in Gibbs potential are equal to zero and a large number of low-symmetry phases converge. Below we discuss several $u_m-T$ diagrams for particular concentrations. Phase diagrams for $x = 1.0$ and $0.8$ presented in Fig.4 are similar to one of the diagrams reported for BT in Ref. 16. In the narrow interval between paraelectric phase \textbf{0} and orthorhombic phase \textbf{3} with in-plane polarization along the basal diagonal of the unit cell there is phase \textbf{2} with the polarization along one of the former cubic axis in plane of the film. Also, in the narrow interval between phase \textbf{1} with the polarization normal to the substrate and phase \textbf{4} with polarization along the space diagonal of the unit cell there is monoclinic phase \textbf{5}. The stability ranges of these intermediate phases (\textbf{2} and \textbf{5}) decreases with the decreasing of Ba content and below $x$ = 0.74 they are not stable any more. As a result, phase transition \textbf{0} $\leftrightarrow$ \textbf{3} and \textbf{1} $\leftrightarrow$ \textbf{4} are the second order, however, there is rather narrow temperature interval where first order \textbf{1} $\leftrightarrow$ \textbf{3} phase transition occurs (see diagram for $x$ = 0.6 in Fig. 4). Below $x$ = 0.4 this first-order boundary disappears and only second-order phase transitions occur for $x = 0.4$ and $0.2$. In this concentration interval phase diagram is similar to that developed for the fourth-order potential. Only three low-symmetry phases (\textbf{1},  \textbf{3} and \textbf{4}) present on the phase diagrams and converge in the multiphase point $T_1$, which steadily decreases with increasing Sr content and obeys the same law as the paraelectric-tetragonal phase transition line in Fig. 1. For $x=0.08$ one more multiphase point $T_2$ appears on the left side phase diagram and two additional phases, namely \textbf{8} and \textbf{16}, are stable for negative misfit strains. Tetragonal phase \textbf{8} $(00\varphi  \text{ } 000)$ is similar to that in bulk ST crystal below 106 K, while phase \textbf{16} $(00\varphi  \text{ } 00p)$ is induced by the mixed order parameter including both octahedral tilting and ionic displacement in the direction normal to the substrate.

The multiphase point $T_2$ moves towards $T_1$ with decreasing $x$ and these points coincide at $x=0.029$. Below this critical concentration, the $T_2$ point appears now on the other side of the diagram at $u_m> 0$, while $T_1$ point is always at $u_m=0$. Three diagrams for $x$=0.025, 0.01, and 0 are presented in Fig. 4. Two low-symmetry phases \textbf{8} $(00\varphi\text{ }  000)$ and \textbf{9} $(\varphi00\text{ } 000)$ converging at $T_1$ correspond to different domains of bulk ST. The transition line between these phases is of first order. In Fig. 4, the paraelectric phase and five low-symmetry phases converge in the $T_2$ point. According to Ref. 14, the existence of such multiphase point $T_2$, where five low-symmetry phases converge, does not contradict the Gibbs phase rule. Below critical concentration $x=0.029$, both $T_1$ and $T_2$ move to higher temperatures with decreasing $x$ and below $T_2$ phase diagrams in Fig. 4 change considerably. Namely, six phases \textbf{0}, \textbf{3}, \textbf{15}, \textbf{32}, \textbf{21} and \textbf{9} converge in the multiphase point $T_2$.

Five phases allowed below $60$ K for $x$ = 0 are shown in Fig. 5c. Phase transitions \textbf{8} $\leftrightarrow$ \textbf{9} and \textbf{18} $\leftrightarrow $\textbf{21} are of the first order, while \textbf{8} $\leftrightarrow$ \textbf{18}, \textbf{9} $\leftrightarrow$ \textbf{21} and \textbf{21} $\leftrightarrow$ \textbf{32} are of the second order. Note all these phases do not converge in one point. There are two close tricritical points so that phase transition line between \textbf{9} and \textbf{18} exists. New phase \textbf{27} appears on the phase diagram at $x$ =0.01 in between \textbf{18} and \textbf{8} phases, while phase \textbf{16} is stable for negative misfit strains (see Fig. 5b).

Near the critical concentration $x$ = 0.029 the phase diagrams are very complicated. Coupling between polarization and structural order parameter in the epitaxial film is modified considerably and new phases that were not present in the bulk material (Fig. 2) appear. Fig. 4 shows overall view and Fig. 5a shows detailed diagram around $u_m$ = 0 for $x$ = 0.025. There are thirteen low-symmetry phases in a film instead of six in a bulk sample. Both rhombohedral phases $R3c$ and $R3m$ allowed in a bulk material are not stable in the film.

Finally, we compare our phase diagram for pure ST film ($x$=0) with those developed by Pertsev $et$ $al$ \cite{c24}. Although phase diagrams have similar overall view for  $ \mid u_m \times 10^3 \mid <10$, some important differences caused by the above discussed changes of the coefficients should be emphasized. As shown in Fig.4, additional phase \textbf{32} exists in between \textbf{21}(FOII) and \textbf{15}(FOIV) phases. We have to emphasize that FOIV phase determined as $\mid p_1 \mid = \mid p_2 \mid$ and $\mid \varphi_1 \mid = \mid \varphi_2 \mid$ (Ref. 24) actually contains two different phases \textbf{15} and \textbf{17} listed in Table II. Note, that ferroelectric (FTI) tetragonal phase \textbf{1} $(000\text{ } 00p)$ predicted in Ref. 27 for negative misfit strains and $T>$200 K, according to our calculations is stable at low temperatures only (see Fig. 4). Also, phase \textbf{15}(FOIV) exists in the limited misfit interval.

It is worth noting that in pure ST many phases are stable in the narrow intervals near $u_m$= 0. Therefore, even weak clamping of single-crystalline plates, usually used for low-temperature measurements, can induce phase transition to one of the low-symmetry phases.

Concluding this section we have to emphasize some limitations of the above developed approach. The phenomenological potential of the solid solution (1) was written through the potentials of its end members ($x=0, 1$). Therefore the validity of the potential (1) is based on validity of the potentials for each member of the solid solution. The method used to develop the potential (1), (2) and (3) is based on the macroscopic theory of elasticity and crystal lattice matching. We assume that solid solution with no ordering in the whole concentration range can be correctly described if ionic substitution induces geometrical distortions, accompanied by the macroscopic elastic deformation \cite{c26}. We also assume that isomorphic Ba/Sr substitution does not change the type of chemical bonding. Therefore, in our model, most (or even all) changes in the solid solution are caused by the elastic interaction.

Actually, currently available potentials for both end-members of BST solid solution require further improvements. Some coefficients for fourth-order potential developed by Uwe \cite{c27} for ST were recently revised \cite{c18}. Although some uncertainty in the values of the coefficients still exists, reasonable agreement with the available experimental data were found \cite{c18}. Further investigations and new perfect experimental data will require extension of the theory to the higher orders \cite{c28}.

Two eight-order potentials are currently available for BT. Li et al \cite{c17} developed a model with only a single temperature-dependent second-order coefficient, while all higher-order coefficients were temperature-independent. This model could successfully reproduce the phase transition temperatures and their dependence on electric fields, as well as the dielectric and piezoelectric constants of bulk BT. Recently, Wang et al \cite{c29} developed potential where forth-order coefficients were also temperature-dependent and successfully explained dielectric properties of BT. Although the latter potential looks much more attractive, we failed to employ it in our model for BST solid solution. Since elastic properties of the potential are very important in our model, we have analyzed the temperature - hydrostatic pressure ($T-p$ ) phase diagram using potential \cite{c29}. We found that the paraelectric-ferroelectric phase transition temperature decreases with increasing pressure, approaches the minimum, then increases, and finally decreases again. Such a behavior contradicts to the experiment \cite{c30}. Although the potential developed by Li \cite{c17} does not reproduces exactly the experimental $T-p$ phase diagram, a disagreement is not so pronounced. Very likely, as was noted in Ref. 31, additional terms should be considered in the potential to describe large deformations.

\section{CONCLUSIONS}

To construct the "concentration-misfit strain" and "temperature- misfit strain" phase diagrams of epitaxial BST-$x$ thin films on (001)-oriented cubic substrates we modified thermodynamic potential for bulk BST-$x$ solid solutions developed in Ref. 14. The thermal expansion, missing in the previous thermodynamic model \cite{c14} was simply introduced in the Helmholtz potential (1) by the shift of the common strain $u$ of the solid solution on the value of the linear thermal expansion. The introduced modification does not change the phase diagram of bulk BST-$x$ solid solution, and allows ones to develop thermodynamic theory for thin films.

In the present work we used eighth-order thermodynamic potential for BT single crystal \cite{c17} and fourth-order potential for ST single crystal \cite{c18} to develop relevant potential of BST-$x$ solid solution, which can be applied at high temperatures.

In order to fit the phase transition lines to experimentally measured phase diagram of BST-$x$ solid solutions \cite{c19}, the coefficients $Q_{11}$ and $Q_{12}$  in six-order term in the potential of BT were changed. Also, the stable thermodynamic states and phase transitions lines in the concentration range $x < 0.04$ were calculated. Five low-symmetry phases are predicted for bulk solid solution in the narrow concentration interval around multiphase point T$_{N2}$ = 48 K, $x$ = 0.029 as shown in Fig. 2.

Performing the necessary calculations, we constructed the room-temperature "concentration- misfit strain" $(x-u_m)$ phase diagram for BST-$x$ thin films epitaxially grown on cubic substrate. The diagram is useful for practical applications in thin-film engineering. Depending on the type of strain imposed by the substrate, epitaxial BST-$x$ thin films can be grown with the polarization either normal or parallel to the substrate. As follows from the phase diagram presented in Fig. 3, ferroelectric state with the polarization parallel to the substrate is only possible is ST films deposited on tensile substrates. Recently \cite{c32}, room-temperature ferroelectricity was observed in ST films on DyScO$_3$ substrates. The tensile lattice mismatch in this heterostructure at room temperature is about 1\%  that is very close to the \textbf{0}-\textbf{3} phase transition line in Fig. 3.

A rich variety of low-symmetry phases induced by octahedral tilting and ionic displacements are allowed near the $N$-phase point  $( u_m= 0, x = 0.029, T = 48$ K). Most of them are stable in the narrow temperature intervals near $u_m$= 0. Great difference between the phase transition sequence in bulk BST-$x$ and epitaxial thin films with the composition around $x$ = 0.029 can be therefore expected. The selection of the substrate and the film composition allow manipulating the strain state in the film to achieve the desirable phase transition sequence.

\begin{acknowledgments}

This study was partially supported by the Russian Foundation for Basic Research Project No. 06-02-16271

\end{acknowledgments}


\end{document}